\documentclass[sigconf]{acmart}
\usepackage{multirow}
\usepackage{makecell}
\AtBeginDocument{%
  }

\begin{document}
\title{Multimodal Generative Retrieval Model with Staged Pretraining for Food Delivery on Meituan}

\author{Boyu Chen}
\affiliation{%
  \institution{Beijing University of Posts and
Telecommunications}
  \city{Beijing}
  \country{China}}
\email{chenbys4@bupt.edu.cn}

\author{Tai Guo}
\author{Weiyu Cui}
\affiliation{%
  \institution{Meituan}
  \city{Beijing}
  \country{China}}
\email{guotai02@meituan.com}
\email{cuiweiyu02@meituan.com}

\author{Yuqing Li}
\author{Xingxing Wang}
\affiliation{%
  \institution{Meituan}
  \city{Beijing}
  \country{China}}
\email{liyuqing16@meituan.com}
\email{wangxingxing04@meituan.com}

\author{Chuan Shi}
\author{Cheng Yang}
\authornote{Corresponding author.}
\affiliation{%
  \institution{Beijing University of Posts and
Telecommunications}
  \city{Beijing}
  \country{China}}
\email{shichuan@bupt.edu.cn}
\email{yangcheng@bupt.edu.cn}

\renewcommand{\shortauthors}{Boyu Chen, Tai Guo, Weiyu Cui, Yuqing Li, Shu Han, Xingxing Wang, Chuan Shi, and Cheng Yang}

\begin{abstract}

Multimodal retrieval models are becoming increasingly important in scenarios such as food delivery, where rich multimodal features can meet diverse user needs and enable precise retrieval. Mainstream approaches typically employ a dual-tower architecture between queries and items, and perform joint optimization of intra-tower and inter-tower tasks. However, we observe that joint optimization often leads to certain modalities dominating the training process, while other modalities are neglected. In addition, inconsistent training speeds across modalities can easily result in the one-epoch problem. To address these challenges, we propose a staged pretraining strategy, which guides the model to focus on specialized tasks at each stage, enabling it to effectively attend to and utilize multimodal features, and allowing flexible control over the training process at each stage to avoid the one-epoch problem. Furthermore, to better utilize the semantic IDs that compress high-dimensional multimodal embeddings, we design both generative and discriminative tasks to help the model understand the associations between SIDs, queries, and item features, thereby improving overall performance. Extensive experiments on large-scale real-world Meituan data demonstrate that our method achieves improvements of 3.80\%, 2.64\%, and 2.17\% on R@5, R@10, and R@20, and 5.10\%, 4.22\%, and 2.09\% on N@5, N@10, and N@20 compared to mainstream baselines. Online A/B testing on the Meituan platform shows that our approach achieves a 1.12\% increase in revenue and a 1.02\% increase in click-through rate, validating the effectiveness and superiority of our method in practical applications.

\end{abstract}



\maketitle

\section{Introduction}
Food delivery search has become an indispensable part of users’ daily lives. Users typically submit textual queries to retrieve a set of relevant items and select desired products for purchase, making the quality of item retrieval a critical factor that directly impacts platform transaction revenue. As user needs become increasingly diverse, relying solely on textual features is no longer sufficient to address all queries. This necessity has led to the emergence of multimodal retrieval models~\cite{yan2025mim,chen2024multi,zhou2023semantic,liu2024alignrec,li2021align,han2022modality}, which integrate and process various types of product features to enable richer and more precise retrieval results. Recent advances in large models have accelerated the development of multimodal retrieval~\cite{sheng2024enhancing,zhang2025notellm,gan2025hcmrm} by enabling the extraction of high-quality feature embeddings, thus providing a strong foundation for item retrieval and matching.

The mainstream framework for multimodal retrieval models is the dual-tower architecture~\cite{yan2025mim,chen2024multi,zhou2023semantic,liu2024alignrec,sheng2024enhancing,zhang2025notellm,gan2025hcmrm,abdool2026applying}, in which the query and item towers respectively encode their feature embeddings. Compared to unimodal retrieval models~\cite{covington2016deep,khattab2020colbert,wu2019neural,wang2018dkn,wu2019npa,wu2025muse,jia2025summa,han2025lemur}, multimodal retrieval models integrate various types of features (e.g., text and images) within the item tower, and employ specialized network modules to process each modality. Moreover, multiple objectives are adopted to optimize the model. To obtain high-quality fused representations, these models typically employ contrastive learning between different modalities of the same sample to ensure consistency (e.g., image2text). Furthermore, to strengthen the association between the two towers, additional contrastive learning tasks are designed between the query and various modality-specific or fused features from the item tower (e.g., query2image, query2text, and query2item), thereby further enhancing the model’s downstream retrieval capability.

\begin{figure}[ht]
    \centering
    \includegraphics[width=\columnwidth]{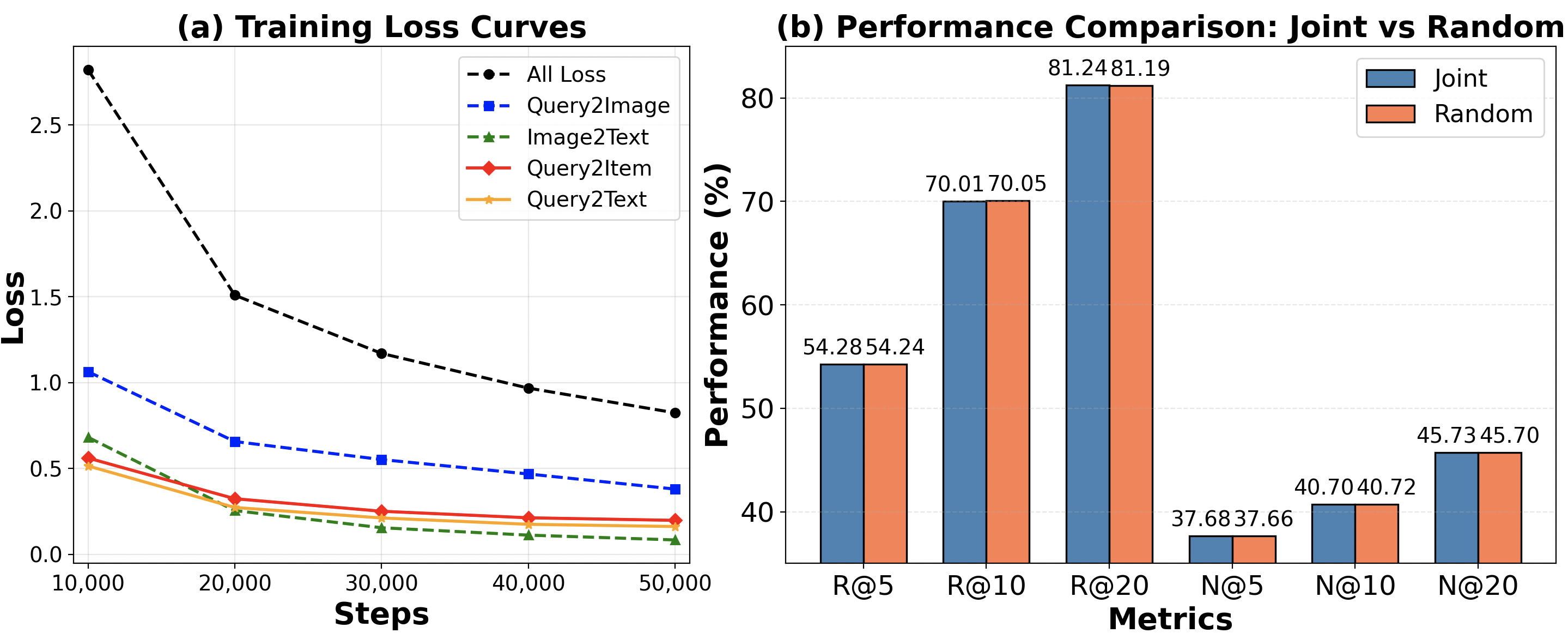}
    \caption{(a) Loss trends of different objectives during joint optimization. (b) Performance comparison between the jointly optimized model and the model with randomly generated image embeddings.}
    \Description{Figure (a) presents line plots illustrating the loss reduction trajectories of various objectives during the pretraining phase. The losses for query2item and query2text exhibit highly consistent trends, while losses for other objectives show limited correlation. Figure (b) further compares the retrieval performance using image embeddings generated by joint training versus randomly generated image embeddings, revealing similar performance and indicating that the model does not effectively attend to or utilize image features.}
    \label{fig:loss}
\end{figure}

Previous multimodal retrieval models typically employ \textbf{joint optimization} of multiple objectives to improve overall performance~\cite{zhou2023semantic,gan2025hcmrm,chen2024multi}. However, as shown in Figure~\ref{fig:loss} (a), our analysis of the loss trajectories during joint training reveals that the reduction in query2item loss closely mirrors that of query2text loss, while losses associated with other modalities exhibit limited correlation. Furthermore, due to differences in learning difficulty across modalities, the loss for image features converges significantly more slowly than that for textual features. This imbalance can lead to the so-called one-epoch problem~\cite{zhang2022towards,hsu2024taming}, where the model overfits features that are easier to learn while under-optimizing more challenging ones, ultimately resulting in degraded overall performance. To further investigate this phenomenon, we conduct an experiment in which image embeddings are replaced with randomly generated vectors. As illustrated in Figure~\ref{fig:loss} (b), the retrieval performance remains nearly unchanged, indicating that the model tends to ignore image features. 

In addition, the introduction of multimodal features in online deployment environments imposes significant storage and computational burdens. To this end, generative retrieval models have emerged, employing residual quantized variational autoencoders (RQ-VAE)~\cite{rajput2023recommender} to discretize high-dimensional embeddings into compact semantic IDs (SIDs), thereby improving retrieval efficiency and feature expressiveness. As newly generated intermediate features, SIDs lack dedicated pretraining in large language models (LLMs), resulting in limited semantic understanding and utilization. Previous methods~\cite{fu2025forge,luo2025qarm,xu2025mmq,wang2025progressive,liang2025tbgrecall,ye2025dual,zhao2025diffugr,ye2025align} typically employ only discriminative losses for downstream fine-tuning, which are insufficient for helping the model understand the associations between SIDs, queries, and item features.

To address these challenges, we propose SMGR, which is a \textbf{m}ultimodal \textbf{g}enerative \textbf{r}etrieval model with \textbf{s}taged
pretraining for food delivery on Meituan. We employ staged pretraining to ensure that the model focuses on specific objectives at each phase, thereby alleviating the issue of other features being neglected due to the dominance of a single modality. Also, this approach effectively avoids the one-epoch problem caused by inconsistent training speeds across modalities. After progressively optimizing high-quality multimodal representations, we leverage RQ-VAE to transform these features into multiple SIDs. Furthermore, we introduce both generative and discriminative tasks for SIDs understanding and downstream adaptation, enabling the model to effectively learn and utilize SIDs. Extensive offline experiments on large-scale real-world Meituan data demonstrate that our method achieves improvements of 3.80\%, 2.64\%, and 2.17\% on R@5, R@10, and R@20, and 5.10\%, 4.22\%, and 2.09\% on N@5, N@10, and N@20 compared to state-of-the-art baselines. In online A/B testing on the Meituan platform, our method also demonstrates significant improvements compared to the online baseline. The contributions of this work are as follows:

\begin{itemize}
    \item We empirically find that, in multimodal retrieval scenarios, joint pretraining may cause the optimization process to be dominated by certain modalities, resulting in the neglect of other modality features. In addition, the joint optimization is prone to the one-epoch problem.
    \item We propose a staged pretraining strategy to address the above limitation. Also, to better utilize the semantic IDs that compress high-dimensional multimodal embeddings, we employ both generative and discriminative tasks to facilitate the model’s understanding, ultimately enhancing overall model performance.
    \item Extensive offline experiments on the Meituan platform demonstrate significant improvements of our proposed method over various baselines. In addition, online A/B testing shows that, our approach achieves a 1.12\% increase in revenue and a 1.02\% increase in click-through rate (CTR), resulting in substantial gains.
\end{itemize}

\section{Related Work}

\subsection{Multimodal Retrieval Model}  
Multimodal dual-tower models~\cite{sheng2024enhancing,zhang2025notellm,yan2025mim,chen2024multi,zhou2023semantic,gan2025hcmrm,liu2024alignrec,li2021align} have demonstrated strong performance in complex retrieval scenarios and have become the mainstream framework for multimodal retrieval tasks. Building upon the traditional dual-tower architecture~\cite{huang2013learning,shen2014latent,covington2016deep}, these models extend the item tower from representing only textual features to incorporating multimodal features such as images, audio, and video. Each modality is encoded using dedicated encoders, including CNNs~\cite{shen2014learning,shen2014latent,huang2013learning}, RNNs, Transformers~\cite{khattab2020colbert,reimers2019sentence}, or ViTs~\cite{yang2022chinese}, and the resulting representations are fused via concatenation~\cite{chen2024multi}, addition~\cite{zhang2025notellm,liu2024alignrec}, multilayer perceptron (MLP)~\cite{sheng2024enhancing}, or attention mechanisms~\cite{zhou2023semantic,gan2025hcmrm} to produce fixed-length multimodal embeddings. During training, contrastive learning is commonly employed for both intra-tower and inter-tower optimization. Intra-tower optimization aims to align representations of the same sample across different modalities to improve feature fusion, while inter-tower optimization focuses on matching relevant query-item pairs to enhance retrieval performance. Some approaches~\cite{zhou2023semantic,gan2025hcmrm,chen2024multi} further design joint optimization strategies to fully exploit multimodal information.

Despite their effectiveness, multimodal dual-tower models face several practical challenges. Modalities with simpler semantics and higher relevance to queries may dominate the optimization process, resulting in underutilization of other modality features. Inconsistent training speeds across modalities can also lead to the one-epoch problem. Additionally, these models require substantial storage and computational resources for deployment. Online inference is often constrained by efficiency, while offline indexing increases storage demands and is susceptible to semantic drift due to data distribution shifts. To maintain performance, periodic model fine-tuning is necessary, but this incurs considerable maintenance costs.

\subsection{Generative Retrieval Model}
Generative retrieval models~\cite{sun2023learning,zhang2023model,wu2024hi,yuan2024generative} have shown significant research value in addressing challenges related to storage, computation, and fine-tuning in recent years. These models leverage the RQ-VAE~\cite{rajput2023recommender} technique from generative recommendation systems, performing residual quantization on original high-dimensional embeddings through multi-layer codebooks to discretize continuous embedding vectors into SIDs. These SIDs can be used directly as item features or as side information to enhance the recall process, thereby improving feature representation quality and retrieval efficiency.

During model training, codebook parameters are jointly optimized using reconstruction loss, codebook loss and commitment loss, enabling the encoder to learn high-quality discrete representations. Specifically, the model maps the original embeddings to positions in multi-layer codebooks, and these positions serve as low-dimensional SIDs for efficient matching and retrieval within the dual-tower architecture. This approach not only significantly reduces storage and computational resource consumption, but also enhances the adaptability of SIDs to dynamic changes in online data distributions, thereby facilitating rapid fine-tuning and efficient deployment.

Subsequent research has further explored the integration of RQ-VAE techniques into multimodal retrieval models~\cite{fu2025forge,luo2025qarm,xu2025mmq,wang2025progressive}, generating SIDs not only for individual modality features but also for their fused representations. These SIDs are partially or fully utilized in downstream retrieval tasks to enhance model performance. However, the proliferation of SIDs introduces new challenges in their selection and utilization. As newly generated features, the effectiveness of multimodal SIDs rely heavily on subsequent fine-tuning, necessitating dedicated training and optimization to fully realize their potential in retrieval tasks. Moreover, efficiently leveraging SIDs during the retrieval process remains a critical issue that significantly impacts the practical performance of these models.

\section{Preliminaries}

In this section, we will introduce the dual-tower dense retrieval framework adopted by the Meituan food delivery retrieval system.

In this framework, each input $x$ represents either a query or an item along with its associated attributes. For each input $x$, we first apply a shared LLM encoder, which processes $x$ according to a predefined prompt template. Specifically, the input text and related attributes are concatenated and fed into the LLM, which tokenizes and encodes the input to produce a sequence of 1024-dimensional vectors. We select the last vector in this sequence (i.e., the EOS position) as the semantic representation of the entire input. Formally, let $x_q$ and $x_i$ denote the input features of a query and an item, respectively. Their semantic representations are given by $e_q = \text{LLM}(x_q)$ and $e_i = \text{LLM}(x_i)$, where $e_q, e_i \in \mathbb{R}^{1024}$.

To meet the requirements of online storage and real-time retrieval, the query tower and item tower independently compress their 1024-dimensional semantic representations using separate MLP, resulting in the final semantic vectors $h_q$ and $h_i$:
\begin{equation}
h_q = \text{MLP}_\mathcal{Q}(e_q), \qquad h_i = \text{MLP}_\mathcal{I}(e_i),
\end{equation}
where $h_q, h_i \in \mathbb{R}^{128}$, and $\text{MLP}_\mathcal{Q}$ and $\text{MLP}_\mathcal{I}$ denote the independent projection networks for the query and item towers, respectively.

During the retrieval phase, the system precomputes and stores the semantic vectors $h_i$ for all candidate items $i \in \mathcal{I}$. For each incoming query $q$, the semantic vector $h_q$ is computed in real time. The relevance between the query and each candidate item is then measured by a similarity function $s(h_q, h_i)$, where cosine similarity is typically used:
\begin{equation}
s(h_q, h_i) = \frac{h_q^\top h_i}{\|h_q\| \|h_i\|}.
\end{equation}

To improve retrieval efficiency, the system employs FAISS~\cite{douze2025faiss} as the approximate nearest neighbor (ANN) search engine. FAISS enables efficient identification of the Top-$K$ items most similar to the query vector from a large-scale item vector database. The final retrieval result for a given query $q$ is formulated as:
\begin{equation}
\mathcal{I}_q = \text{Top-}K(\{i : s(h_q, h_i) \mid i \in \mathcal{I}\}).
\end{equation}

Considering the potential semantic gap between different features, we adopt contrastive learning to reduce the distribution discrepancy and obtain high-quality semantic embeddings. The core idea of contrastive learning is to optimize the model such that the semantic representations of related samples are closer in the embedding space, while those of unrelated samples are effectively separated. Here, related samples typically refer to positive samples matching the ground truth, and unrelated samples are negative samples obtained through random sampling or other strategies. For unified reference in subsequent method sections, we define the basic contrastive learning loss as follows:
\begin{equation}
\mathcal{L}_{\text{InfoNCE}}(x, x^+, \{x^-\}) = -\log \frac{\exp\left(\frac{s(x, x^+)}{\tau}\right)}{\exp\left(\frac{s(x, x^+)}{\tau}\right) + \sum_{x^-} \exp\left(\frac{s(x, x^-)}{\tau}\right)},
\end{equation}
where $x$ denotes the input sample, $x^+$ denotes the positive sample semantically related to $x$, $\{x^-\}$ denotes the set of negative samples unrelated to $x$, $s(\cdot, \cdot)$ is a similarity function (e.g., cosine similarity), and $\tau$ is a temperature coefficient that controls the sharpness of the similarity distribution.

\begin{figure*}[ht]
  \centering
  \includegraphics[width=0.8\textwidth,height=0.34\textheight]{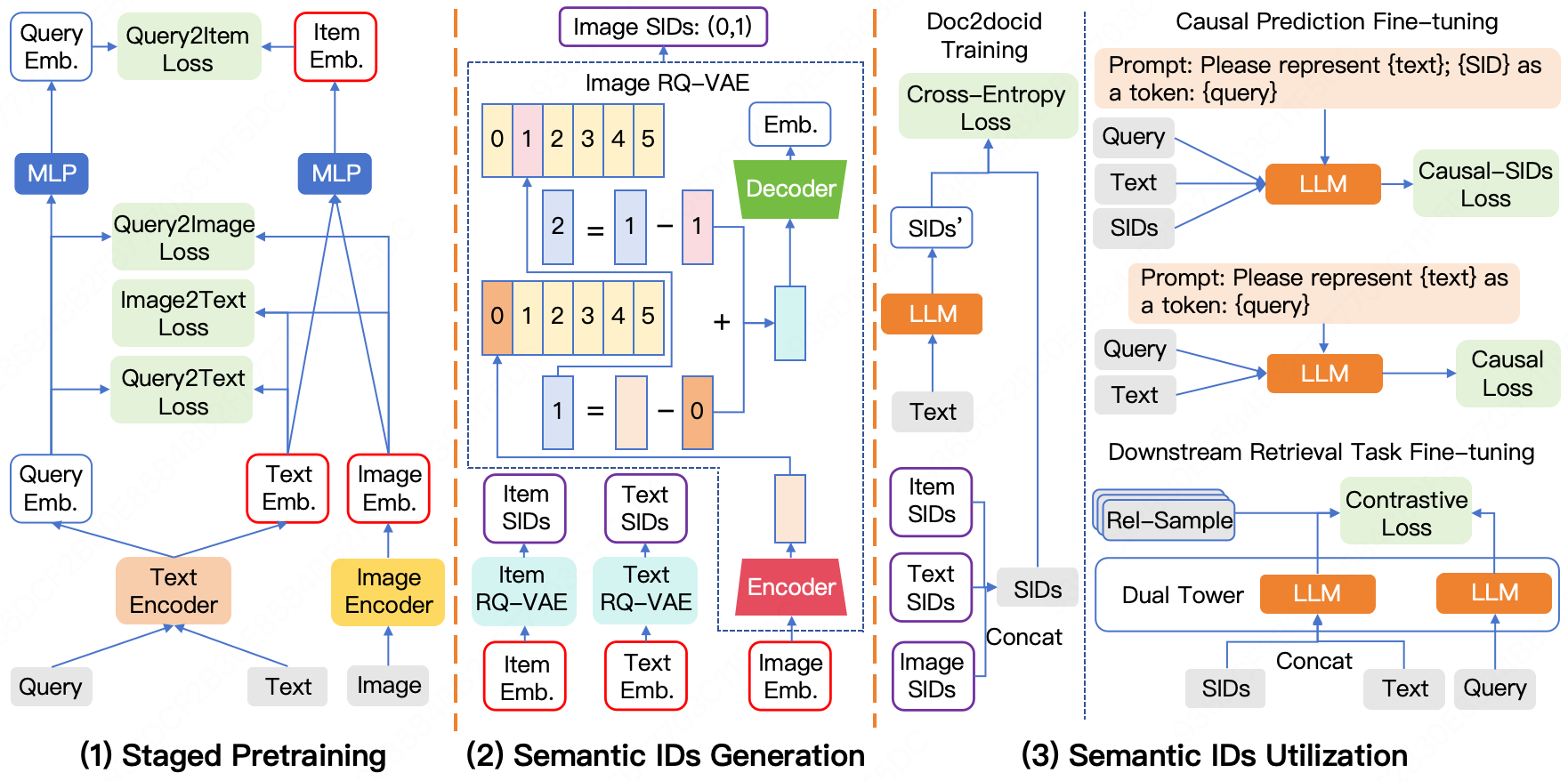}
  \caption{The overall framework of our proposed model with three principal components. 1) Staged Pretraining: High-quality multimodal embeddings are obtained through staged pretraining. 2) Semantic IDs Generation: High-dimensional embeddings are transformed into discrete SIDs to alleviate deployment burden. 3) Semantic IDs Utilization: The model is fine-tuned to adapt to SIDs, thereby enhancing downstream task performance.}
  \Description{The overall framework of our proposed model with three main stages. 1) Node Retrieval Stage: High-quality multimodal embeddings are obtained through staged pretraining. 2) Semantic IDs Generation: High-dimensional embeddings are transformed into discrete SIDs to alleviate deployment burden. 3) Semantic IDs Utilization: The model is fine-tuned to adapt to SIDs, thereby enhancing downstream task performance.}
  \label{fig:model}
\end{figure*}

\section{Methodology}

Our method consists of three principal components: staged pretraining for embedding learning, semantic IDs generation via residual quantization, and semantic IDs utilization for downstream retrieval and adaptation. The overall workflow is illustrated in Figure~\ref{fig:model}.

\subsection{Staged Pretraining}

Traditional dual-tower models, which only perform contrastive learning between textual queries and items, are insufficient for real-world multimodal scenarios. In Meituan’s food delivery retrieval system, the query tower typically contains only textual query features, while the item tower encompasses both textual descriptions and images of goods. Thus, we define the input as $(q, (i_{\text{text}}, i_{\text{image}}))$, where $q \in \mathcal{Q}$ denotes the query text, $(i_{\text{text}}, i_{\text{image}}) \in \mathcal{I}$ represents the textual and image features of an item, $\mathcal{Q}$ is the set of queries, and $\mathcal{I}$ is the set of items.

To reduce the multimodal gap within the item tower, we will perform contrastive learning between the textual and image features. Specifically, we pull together the representations of text and image features from the same sample and push apart those from different samples, yielding high-quality multimodal representations. This task is defined as image2text, with the loss function:
\begin{equation}
\mathcal{L}_{\text{image2text}} = \mathcal{L}_{\text{InfoNCE}}(h_{\text{image}}, h_{\text{text}}^+, \{h_{\text{text}}^-\}),
\end{equation}
where negative samples are randomly sampled, $h_{\text{text}} = \text{LLM}(i_{\text{text}})$ and $h_{\text{image}} = \text{ViT}(i_{\text{image}})$, with ViT is a vision transformer used to obtain semantic embeddings from images.

In downstream retrieval tasks, although the query ultimately retrieves the fused multimodal item features, it is critical to enhance the model’s understanding and responsiveness to each modality by introducing additional contrastive learning objectives between the query and various item modalities. Specifically, we define query2item, query2text, and query2image tasks, with the following loss functions:
\begin{equation}
\left\{
\begin{aligned}
&\mathcal{L}_{\text{query2item}} = \mathcal{L}_{\text{InfoNCE}}(\text{MLP}_\mathcal{Q}(h_q), \text{MLP}_\mathcal{I}(h_{\text{item}}^+), \{\text{MLP}_\mathcal{I}(h_{\text{item}}^-)\}), \\
&\mathcal{L}_{\text{query2text}} = \mathcal{L}_{\text{InfoNCE}}(h_q, h_{\text{text}}^+, \{h_{\text{text}}^-\}), \\
&\mathcal{L}_{\text{query2image}} = \mathcal{L}_{\text{InfoNCE}}(h_q, h_{\text{image}}^+, \{h_{\text{image}}^-\}),
\end{aligned}
\right.
\end{equation}
where $h_q$, $h_{\text{text}}$, and $h_{\text{image}}$ are query, text, and image embeddings encoded by the LLM, and $h_{\text{item}} = h_{\text{text}} + h_{\text{image}}$ is the fused item embedding. $\mathrm{MLP}_\mathcal{Q}$ and $\mathrm{MLP}_\mathcal{I}$ are projection layers for dimensionality reduction.

A common approach is joint optimization~\cite{zhou2023semantic}, where all four losses are summed for overall model training:
\begin{equation}
\mathcal{L}_{\text{mtrain}} = \mathcal{L}_{\text{image2text}} + \mathcal{L}_{\text{query2text}} + \mathcal{L}_{\text{query2image}} + \mathcal{L}_{\text{query2item}}.
\end{equation}
Since joint optimization often leads to the neglect of certain modality features and the one-epoch problem, significantly compromising model performance. To address these issues, we propose a staged pretraining strategy. In the first phase, the text encoder is optimized by aligning query and text features to enhance semantic consistency. The second phase applies contrastive learning between image and text features, strengthening multimodal representations in the item tower. The third phase focuses on aligning queries with image features to improve the query tower’s understanding of visual information. Finally, end-to-end optimization is performed for query-item matching to maximize retrieval performance.

This staged training strategy enables progressive optimization from simple to complex objectives, ensures focused learning at each stage, and effectively alleviates the one-epoch problem, thereby significantly enhancing the overall performance of multimodal retrieval models.

\subsection{Semantic IDs Generation}

After staged pretraining, we obtain high-quality multimodal embeddings and their corresponding pretrained encoders. However, resource constraints in online deployment environments make it challenging to directly utilize multimodal embeddings and to fine-tune multimodal models. To address these issues, we draw inspiration from RQ-VAE~\cite{rajput2023recommender} techniques in generative recommendation systems. We encode multimodal feature embeddings into SIDs that retain semantic information and concatenate them as side information to the item tower's text features for downstream fine-tuning. This approach not only significantly reduces storage, computation, and inference burdens, but also enhances the text encoder's understanding and utilization of multimodal features. More details of RQ-VAE are provided in Appendix~\ref{appendix:RQ-VAE}.

We construct three independent three-layer RQ-VAE codebooks for text, image, and fused modalities, denoted as $C_{\text{text}}$, $C_{\text{image}}$, and $C_{\text{item}}$, respectively. The encoding process for each modality can be uniformly represented as
\begin{equation}
g_{*}(i_{*}) = f_{\text{RQ-VAE}}(h_*) = [\mathrm{SIDs}_{*}^{1}, \mathrm{SIDs}_{*}^{2}, \mathrm{SIDs}_{*}^{3}],
\label{eq:rqvae_sid}
\end{equation}
where $*$ stands for text, image, or item. Here, $h_*$ denotes the high-dimensional embedding of the corresponding modality, and $f_{\text{RQ-VAE}}(\cdot)$ represents the RQ-VAE encoding function. $\mathrm{SIDs}_{*}^{k}$ is the discrete semantic identifier produced by the $k$-th layer codebook for the corresponding modality. This hierarchical residual quantization compresses high-dimensional embeddings into compact and semantically meaningful discrete sequences, facilitating efficient indexing and retrieval.

During practical deployment, the SIDs from the fused, image, and text modalities are concatenated with the item's textual features for the item tower's text encoder, enabling embedding and downstream retrieval fine-tuning. This design not only preserves the semantic information of multimodal features, but also effectively reduces the system burden of storage, computation, and inference, and alleviates performance degradation caused by feature distribution shift. We uniformly define the $SIDs$ as:
\begin{equation}
SIDs = [g_{\text{text}}(i_{\text{text}}),\; g_{\text{image}}(i_{\text{image}}),\; g_{\text{item}}(i_{\text{item}})].
\label{eq:sid_def}
\end{equation}
The final item embedding can be concisely expressed as:
\begin{equation}
h_i = \mathrm{MLP}_\mathcal{I}\left(\mathrm{LLM}\left(\mathrm{concat}(i_{\text{text}},\ SIDs)\right)\right)
\label{eq:item_embedding},
\end{equation}
where $\mathrm{concat}(\cdot)$ denotes the operation of merging these multimodal features into a unified input.

\subsection{Semantic IDs Utilization}
Now we have replaced high-dimensional embeddings with SIDs as side information, effectively alleviating storage and inference burdens. However, the model still faces challenges in understanding and utilizing the newly introduced SIDs features. To address this issue, we design a series of targeted tasks, detailed as follows:

\textbf{Doc2docid Training.} The newly introduced SIDs, as special tokens, are not present in the original vocabulary of the LLM, making it difficult for the model to directly capture their semantic meaning. Thus, we expand the LLM's vocabulary to include all SIDs and design a supervised task to enhance its understanding. Specifically, given textual features, the model is required to predict SIDs for three modalities, enabling the LLM to learn the mapping between text and SIDs. This task not only helps the LLM comprehend the actual semantics represented by the SIDs, but also strengthens the semantic association between SIDs and textual features, thereby improving the quality of item-side feature embeddings and retrieval performance. The loss function is defined as:
\begin{equation}
\mathcal{L}_{\text{doc2docid}} = \mathbb{E}_{(i_{\text{text}},\, SIDs)}\left[\mathrm{CrossEntropy}\left(\mathrm{LLM}(i_{\text{text}}),\, SIDs\right)\right],
\end{equation}
where $\mathrm{CrossEntropy}(\cdot,\cdot)$ denotes the cross-entropy loss, which measures the accuracy of the LLM's SIDs prediction and encourages semantic alignment between SIDs and text. 

\textbf{Causal Prediction Fine-tuning.}
To further enhance the model’s ability to capture associations between item-side and query-side features, we design a causal prediction fine-tuning task, where the model predicts the next token based on the previous context. Considering that SIDs are newly introduced features and the LLM has limited prior understanding of them, directly fine-tuning with SIDs can be challenging. Therefore, we also design an auxiliary task without SIDs to facilitate the learning process. The specific loss functions are defined as follows:
\begin{equation}
\left\{
\begin{aligned}
&\mathcal{L}_{\text{causal}} = \mathbb{E}_{(i_{\text{text}}, q)}\left[-\sum_{t=1}^{T-1} \log P(q_{t+1} \mid i_{\text{text}}, q_{1:t})\right], \\
&\mathcal{L}_{\text{causal\_SIDs}} = \mathbb{E}_{(i_{\text{text}}, SIDs, q)}\left[-\sum_{t=1}^{T-1} \log P(q_{t+1} \mid i_{\text{text}}, SIDs, q_{1:t})\right],
\end{aligned}
\right.
\end{equation}
where $T$ denotes the length of the query sequence, $t$ is the current time step, and $P(q_{t+1} \mid \cdot)$ represents the probability that the model predicts the next query token in a causal manner given the previous context. $\mathcal{L}_{\text{causal}}$ performs causal prediction based only on item textual features, while $\mathcal{L}_{\text{causal\_SIDs}}$ incorporates both item textual features and SIDs.

\textbf{Downstream Retrieval Task Fine-tuning:}
For the downstream query-to-item retrieval task, we employ the loss function $\mathcal{L}_{\text{query2item}}$ to fine-tune the model via contrastive learning, thereby enhancing its ability to capture associations between dual-tower features. In addition to random negative sampling, we further introduce more challenging negative samples. Specifically, after the retrieval stage, the online framework utilizes a relevance model to filter candidate items; among these, some items are highly relevant to the query but were not actually clicked by the user. Such items contain implicit user preference information and are more difficult for the model to learn. By combining random negative sampling with the mining of more challenging negatives, we derive two contrastive learning loss functions to optimize the model's downstream retrieval performance.

In this component, we focus on the newly introduced SIDs features. We first employ doc2docid training to facilitate the model’s understanding of these features. Subsequently, the model is further optimized using a combined loss from causal prediction fine-tuning and downstream retrieval task fine-tuning. This training strategy enables the model to establish and strengthen associations between query and item features, while also deepening its understanding of the SIDs features, thereby significantly enhancing its representational capacity and retrieval performance in retrieval tasks.

\begin{table*}[t]
\centering
\caption{Performance (\%) comparison across different query types and datasets. R@K and N@K denote Recall@K and NDCG@K, respectively. * indicates statistically significant improvements (t-test, $p < 0.05$) over the best baseline.}
\label{tab:main results}
\resizebox{\textwidth}{!}{
\begin{tabular}{c|c|cccccc|cccccc}
\toprule
\multirow{2}{*}{\textbf{Query Type}} & \multirow{2}{*}{\textbf{Model}} & \multicolumn{6}{c|}{\textbf{MT-Popular Cities}} & \multicolumn{6}{c}{\textbf{MT-Other Cities}} \\
\cmidrule(lr){3-8} \cmidrule(lr){9-14}
& & \textbf{R@5} & \textbf{N@5} & \textbf{R@10} & \textbf{N@10} & \textbf{R@20} & \textbf{N@20} & \textbf{R@5} & \textbf{N@5} & \textbf{R@10} & \textbf{N@10} & \textbf{R@20} & \textbf{N@20} \\
\midrule
\midrule
\multirow{5}{*}{\textbf{All Queries}} 
& Qwen3-DualTower & 53.48 & 36.98 & 68.99 & 40.17 & 80.46 & 45.34 & 55.77 & 39.32 & 70.35 & 44.07 & 81.98 & 47.77 \\
& Joint-Que2search & 53.78 & 37.22 & 69.46 & 40.26 & 80.73 & 45.41 & 55.89 & 39.45 & 70.42 & 44.17 & 82.04 & 47.82 \\
& TIGER & 53.82 & 37.30 & 69.49 & 40.31 & 80.74 & 45.45 & 55.97 & 39.51 & 70.49 & 44.22 & 82.11 & 47.87 \\
& MTIGER & 56.01 & 39.58 & 71.18 & 42.08 & 82.42 & 47.17 & 57.95 & 41.53 & 72.44 & 46.45 & 83.56 & 49.98 \\
& SMGR & \textbf{58.19$^{*}$} & \textbf{41.65$^{*}$} & \textbf{72.92$^{*}$} & \textbf{44.02$^{*}$} & \textbf{84.10$^{*}$} & \textbf{48.16$^{*}$} & \textbf{60.10$^{*}$} & \textbf{43.59$^{*}$} & \textbf{74.50$^{*}$} & \textbf{48.23$^{*}$} & \textbf{85.48$^{*}$} & \textbf{51.02$^{*}$} \\
\midrule
\midrule
\multirow{5}{*}{\textbf{High-Freq}} 
& Qwen3-DualTower & 44.68 & 30.27 & 61.15 & 34.38 & 75.74 & 39.81 & 47.23 & 32.34 & 63.78 & 36.93 & 77.46 & 42.44 \\
& Joint-Que2search & 45.05 & 30.66 & 61.50 & 34.52 & 75.87 & 39.91 & 47.68 & 32.67 & 63.93 & 37.04 & 77.54 & 42.58 \\
& TIGER & 45.38 & 31.02 & 61.79 & 34.88 & 75.99 & 40.14 & 47.87 & 32.85 & 64.04 & 37.18 & 77.65 & 42.67 \\
& MTIGER & 47.91 & 32.98 & 63.68 & 36.45 & 77.58 & 41.20 & 50.05 & 35.04 & 66.36 & 39.52 & 78.92 & 44.28 \\
& SMGR & \textbf{51.12$^{*}$} & \textbf{35.09$^{*}$} & \textbf{66.74$^{*}$} & \textbf{38.42$^{*}$} & \textbf{79.50$^{*}$} & \textbf{42.39$^{*}$} & \textbf{53.42$^{*}$} & \textbf{37.13$^{*}$} & \textbf{69.49$^{*}$} & \textbf{41.44$^{*}$} & \textbf{80.78$^{*}$} & \textbf{45.91$^{*}$} \\
\bottomrule
\end{tabular}
}
\end{table*}

\section{Experiments}
We conduct extensive experiments to answer the following research questions (RQs):
\textbf{RQ1:} How does SMGR perform compared to state-of-the-art vector retrieval baselines?
\textbf{RQ2:} How does staged pretraining contribute to the quality of image encoder embeddings?
\textbf{RQ3:} How does the use of SIDs improve the utilization and representation of multimodal features?
\textbf{RQ4:} How important is the adaptation of SIDs to downstream tasks for overall model effectiveness?
\textbf{RQ5:} How does SMGR perform in real-world online retrieval scenario with industry-standard metrics?
\textbf{RQ6:} How do case examples illustrate the advantages of SMGR over the baseline in real-world scenarios?

\subsection{Experimental Setup}

\textbf{Datasets.} 
We construct our experimental dataset using real-world data from the Meituan platform, one of the largest food delivery platforms in China. Specifically, user interaction and item data spanning one week are selected for training and validation, comprising approximately 32 million samples. Each sample includes dish ID (SPU\_id), dish name (SPU\_name), restaurant name (POI\_name), item keywords (keywords), region information (geo\_hash), and multimodal semantic IDs (item\_SID, image\_SID, text\_SID) generated by our method. To better adapt the model to real-world user preferences, we collect hard negative samples for each query, defined as items deemed highly relevant by our relevance model but not actually selected by users; these hard negatives possess the same feature structure as the standard training samples.

For evaluation, we use real click records from the subsequent two days and partition the evaluation set by city popularity into \textit{MT-Popular Cities} (e.g., Beijing, Shanghai, Guangzhou) and \textit{MT-Other Cities}. Each evaluation set contains a candidate pool of 5.8 million items and 2 million user clicks. To further assess performance under more challenging conditions, we construct high-frequency subsets by selecting queries with frequency greater than 100 from both partitions.

\textbf{Baselines.} To comprehensively assess the effectiveness of our proposed framework, we conduct comparative experiments with several strong baseline methods across various datasets:
\begin{itemize}
    \item \textbf{Qwen3-DualTower}: A dual-tower retrieval model based on the Facebook EBR architecture~\cite{huang2020facebookebr}, where the query and item encoders are replaced with full Qwen3~\cite{yang2025qwen3} models as text encoders, providing robust feature encoding, instruction-following, and side information modeling capabilities.
    \item \textbf{Joint-Que2search}: This method employs joint training optimization within the Que2search framework~\cite{liu2021que2search}, obtaining image embeddings through joint training and fusing them with text embeddings via an MLP for subsequent retrieval.
    \item \textbf{TIGER}: Utilizes RQ-VAE~\cite{rajput2023recommender} technology to encode text embeddings into textual SIDs, which are then used alongside textual features for retrieval.
    \item \textbf{MTIGER}: Follows the TIGER paradigm by quantizing text, image, and fused multimodal embeddings into SIDs, concatenating them with item-side textual features, and subsequently embedding them for retrieval.
\end{itemize}

\textbf{Implementation Details.} 
For fair and consistent evaluation, all experiments use Qwen3-0.6B~\cite{yang2025qwen3} as the text encoder and cn-CLIP-ViT-h-14~\cite{yang2022chinese} as the image encoder. We use three-layer MLPs to project features from $1024$ to $128$ dimensions for efficient deployment. Codebooks are initialized via K-means and trained with RQ-VAE, each with three quantization layers. The hidden layers adopt a pyramidal structure: $[1024, 768, 512, 256]$ for text/image and $[128, 64, 32]$ for multimodal features. Experiments are conducted on eight NVIDIA A100 GPUs (80GB each). The batch size is $8$ for staged pretraining and $16$ for downstream fine-tuning (with gradient accumulation step $8$). The contrastive learning temperature is fixed at $0.05$. Additional hyperparameter details are provided in the Appendix~\ref{appendix:hyperparameter}.

\textbf{Evaluation Metrics.} For retrieval evaluation, FAISS~\cite{douze2025faiss} is employed for efficient ANN search and regional constraints are applied via $geo\_hash$. Recall@K and NDCG@K ($K \in \{5, 10, 20\}$) are used as primary metrics. Each experiment is repeated ten times, and statistical significance is assessed using a paired t-test; results with $p < 0.05$ are marked with $^{*}$. More details of evaluation protocol are provided in Appendix~\ref{appendix:evaluation}.

\subsection{Main Results (RQ1)}
As shown in Table~\ref{tab:main results}, our method consistently outperforms all baseline methods on both datasets. Compared with the best baseline, MTIGER, our method achieves average improvements of 3.80\%, 2.64\%, and 2.17\% on the R@5, R@10, and R@20 metrics, respectively, and average improvements of 5.10\%, 4.22\%, and 2.09\% on the N@5, N@10, and N@20 metrics, respectively. Furthermore, on the dataset composed of high-frequency queries, the increased number of candidate samples raises the difficulty of the retrieval task, leading to decreases in R@K and N@K metrics for all methods. Nevertheless, our method still maintains optimal performance, with average improvements over MTIGER of 6.70\%, 4.81\%, and 2.47\% on R@5, R@10, and R@20, and 6.40\%, 5.40\%, and 2.89\% on N@5, N@10, and N@20, respectively. These improvements are greater than those observed on the complete dataset, indicating that our method can consistently achieve superior performance even in more challenging scenarios, and the improvements are more pronounced, allowing for better adaptation to complex tasks. 

We also observe that the Joint-Que2search model, which directly utilizes multimodal embeddings for retrieval, achieves performance comparable to the Qwen3-DualTower model that does not incorporate multimodal features, and performs worse than the TIGER model, which transforms text embeddings into textual SIDs. In contrast, methods that leverage multimodal SIDs, such as MTIGER and SMGR, demonstrate significantly superior performance. These results highlight the necessity of not only incorporating multimodal features but also transforming them into SIDs for effective retrieval.

\begin{table}[ht]
\centering
\caption{Performance (\%) comparison of different multimodal feature learning methods. ``Joint'' refers to the model variant utilizing jointly trained image embeddings, while ``Random'' denotes the variant with randomly generated image embeddings. ``Order'' represents model variants employing staged training for image embedding generation, which are further divided into six types based on the training order; among them, ``Order6'' corresponds to the training sequence adopted in our proposed method. ``*'' indicates statistically significant improvements (t-test, p < 0.05) over the best variant. Further details are provided in Appendix~\ref{appendix:ablation_variants}.}
\label{tab:rq2 type2}
\resizebox{\columnwidth}{!}{
\begin{tabular}{c|c|cccccc}
\toprule
\textbf{Dataset} & \textbf{Variant} & \textbf{R@5} & \textbf{N@5} & \textbf{R@10} & \textbf{N@10} & \textbf{R@20} & \textbf{N@20} \\
\midrule
\midrule
\multirow{8}{*}{\textbf{\makecell{MT-Popular\\Cities}}} 
& Joint & 54.28 & 37.68 & 70.01 & 40.70 & 81.24 & 45.73 \\
& Random & 54.24 & 37.66 & 70.05 & 40.72 & 81.19 & 45.70 \\
& Order1 & 54.38 & 37.79 & 70.17 & 40.81 & 81.29 & 45.77 \\
& Order2 & 54.63 & 38.06 & 70.32 & 40.99 & 81.50 & 46.02 \\
& Order3 & 54.67 & 38.12 & 70.35 & 41.03 & 81.51 & 46.07 \\
& Order4 & 54.76 & 38.27 & 70.40 & 41.19 & 81.60 & 46.27 \\
& Order5 & 54.82 & 38.32 & 70.45 & 41.25 & 81.69 & 46.32 \\
& Order6 (ours) & \textbf{54.92$^{*}$} & \textbf{38.45$^{*}$} & \textbf{70.52$^{*}$} & \textbf{41.37$^{*}$} & \textbf{81.77$^{*}$} & \textbf{46.48$^{*}$} \\
\midrule
\midrule
\multirow{8}{*}{\textbf{\makecell{MT-Other\\Cities}}} 
& Joint & 56.28 & 39.91 & 70.79 & 44.62 & 82.34 & 48.21 \\
& Random & 56.22 & 39.86 & 70.71 & 44.56 & 82.29 & 48.17 \\
& Order1 & 56.36 & 39.99 & 70.82 & 44.66 & 82.38 & 48.27 \\
& Order2 & 56.53 & 40.19 & 70.96 & 44.87 & 82.55 & 48.47 \\
& Order3 & 56.58 & 40.23 & 70.99 & 44.89 & 82.56 & 48.50 \\
& Order4 & 56.66 & 40.28 & 71.07 & 44.92 & 82.59 & 48.60 \\
& Order5 & 56.71 & 40.37 & 71.11 & 45.02 & 82.62 & 48.67 \\
& Order6 (ours) & \textbf{56.84$^{*}$} & \textbf{40.49$^{*}$} & \textbf{71.23$^{*}$} & \textbf{45.13$^{*}$} & \textbf{82.73$^{*}$} & \textbf{48.81$^{*}$} \\
\bottomrule
\end{tabular}
}
\end{table}

\subsection{Training Strategies (RQ2)}
As shown in Table~\ref{tab:rq2 type2}, the ``Joint'' variant achieves performance that is nearly identical to the ``Random'' variant, indicating that the model trained with joint optimization essentially ignores image features and fails to effectively utilize visual information. In contrast, our proposed method achieves consistent improvements on both datasets, with average gains of 1.09\%, 0.68\%, and 0.56\% on R@5, R@10, and R@20, and 1.75\%, 1.39\%, and 1.44\% on N@5, N@10, and N@20, respectively, significantly outperforming the ``Joint'' variant. These results indicate that staged pretraining can effectively mitigate the underutilization of image features, thereby enhancing the overall performance of the model.

Furthermore, we conduct ablation studies on the order of staged pretraining. Our approach consistently achieves the best performance across both datasets, with average improvements of 0.21\%, 0.13\%, and 0.12\% on R@5, R@10, and R@20, and 0.32\%, 0.27\%, and 0.32\% on N@5, N@10, and N@20, respectively, compared to the next best order (``Order5''). Additionally, in inter-tower contrastive learning, a design from easier to more difficult tasks proves to be necessary. Specifically, learning query2text before query2image effectively improves model performance (as evidenced by ``Order5'' outperforming ``Order4'', ``Order3'' outperforming ``Order2'', and ``Order6'' outperforming ``Order1''). Notably, all staged pretraining variants outperform the ``Joint'' variant regardless of the training order, further confirming the necessity and effectiveness of the staged pretraining strategy.

\begin{table}[ht]
\centering
\caption{Performance (\%) comparison of different multimodal feature utilization methods. Specifically, item-only, image-text, item-image, and item-text refer to variants that utilize only the SIDs corresponding to the respective modality. * indicates statistically significant improvements (t-test, p < 0.05) over the best variant.}
\label{tab:rq3 type2}
\resizebox{\columnwidth}{!}{
\begin{tabular}{c|c|cccccc}
\toprule
\textbf{Dataset} & \textbf{Variant} & \textbf{R@5} & \textbf{N@5} & \textbf{R@10} & \textbf{N@10} & \textbf{R@20} & \textbf{N@20} \\
\midrule
\midrule
\multirow{6}{*}{\textbf{\makecell{MT-Popular\\Cities}}} 
& Order6 & 54.92 & 38.45 & 70.52 & 41.37 & 81.77 & 46.48 \\
& item-only & 55.98 & 39.41 & 71.21 & 42.03 & 82.37 & 46.97 \\
& image-text & 56.52 & 39.98 & 71.70 & 42.42 & 82.78 & 47.21 \\
& item-text & 57.29 & 40.76 & 72.27 & 43.13 & 83.25 & 47.66 \\
& item-image & 57.59 & 40.99 & 72.39 & 43.35 & 83.49 & 47.88 \\
& SMGR & \textbf{58.19$^{*}$} & \textbf{41.65$^{*}$} & \textbf{72.92$^{*}$} & \textbf{44.02$^{*}$} & \textbf{84.10$^{*}$} & \textbf{48.16$^{*}$} \\
\midrule
\midrule
\multirow{6}{*}{\textbf{\makecell{MT-Other\\Cities}}} 
& Order6 & 56.84 & 40.49 & 71.23 & 45.13 & 82.73 & 48.81 \\
& item-only & 57.78 & 41.36 & 72.07 & 46.01 & 83.56 & 49.34 \\
& image-text & 58.34 & 41.92 & 72.71 & 46.63 & 84.08 & 49.81 \\
& item-text & 59.19 & 42.69 & 73.42 & 47.32 & 84.78 & 50.36 \\
& item-image & 59.43 & 42.83 & 73.74 & 47.56 & 84.94 & 50.48 \\
& SMGR & \textbf{60.10$^{*}$} & \textbf{43.59$^{*}$} & \textbf{74.50$^{*}$} & \textbf{48.23$^{*}$} & \textbf{85.48$^{*}$} & \textbf{51.02$^{*}$} \\
\bottomrule
\end{tabular}
}
\end{table}

\subsection{Utilization of Multimodal Features (RQ3)}
We conduct comparative experiments to investigate whether directly using image embeddings or converting them into SIDs and incorporating them as side information for item features would yield better performance. As shown in Table ~\ref{tab:rq3 type2}, compared to ``Order6'', which directly uses image embeddings, our method achieves average improvements of 5.84\%, 3.99\%, and 3.09\% on R@5, R@10, and R@20, and 7.99\%, 6.64\%, and 4.07\% on N@5, N@10, and N@20 across both datasets.

Additionally, we perform ablation studies on the SIDs used. The results show that using only the item SIDs yields the worst performance, mainly because the compressed information it contains is insufficient to support inference. Similarly, the image-text method lacks the associative information provided by item SID, resulting in degraded model performance. Although item-image and item-text methods achieve relatively good results, their performance is suboptimal due to the absence of SIDs information from the other modality. Notably, all these methods outperform the approach of directly using image embeddings, demonstrating the necessity of converting features into SIDs before use. Moreover, the results indicate that jointly utilizing all SIDs is essential for achieving optimal performance.

\subsection{Adaptation to Downstream Tasks (RQ4)}
SIDs are difficult to apply directly and require appropriate fine-tuning tasks to better adapt to downstream applications. As shown in Table~\ref{tab:rq4 type2}, compared to the variant that only employs a doc2docid training task, all variants that incorporate causal prediction fine-tuning achieve consistent performance improvements. Specifically, our approach achieves average improvements of 3.00\%, 2.13\%, and 1.73\% on R@5, R@10, and R@20, and 4.16\%, 3.32\%, and 1.55\% on N@5, N@10, and N@20 across both datasets. Furthermore, we observe that, compared to methods that use only $\mathcal{L}_{\text{causal}}$ or $\mathcal{L}_{\text{causal\_SIDs}}$, using both losses together facilitates the model’s adaptation to newly introduced features, thereby achieving optimal performance.

\begin{table}[ht]
\centering
\caption{Performance (\%) comparison of methods employing different fine-tuning tasks for downstream adaptation. * indicates statistically significant improvements (t-test, p < 0.05) over the best variant.}
\label{tab:rq4 type2}
\resizebox{\columnwidth}{!}{
\begin{tabular}{c|c|cccccc}
\toprule
\textbf{Dataset} & \textbf{Variant} & \textbf{R@5} & \textbf{N@5} & \textbf{R@10} & \textbf{N@10} & \textbf{R@20} & \textbf{N@20} \\
\midrule
\midrule
\multirow{4}{*}{\textbf{\makecell{MT-Popular\\Cities}}} 
& $\mathcal{L}_{doc2docid}$ & 56.44 & 39.95 & 71.53 & 42.48 & 82.78 & 47.37 \\
& $\mathcal{L}_{doc2docid}+\mathcal{L}_{causal}$ & 57.53 & 40.92 & 72.36 & 43.44 & 83.59 & 47.87 \\
& $\mathcal{L}_{doc2docid}+\mathcal{L}_{causal\_SIDs}$ & 57.33 & 40.70 & 72.18 & 43.25 & 83.37 & 47.69 \\
& SMGR & \textbf{58.19$^{*}$} & \textbf{41.65$^{*}$} & \textbf{72.92$^{*}$} & \textbf{44.02$^{*}$} & \textbf{84.10$^{*}$} & \textbf{48.16$^{*}$} \\
\midrule
\midrule
\multirow{4}{*}{\textbf{\makecell{MT-Other\\Cities}}} 
& $\mathcal{L}_{doc2docid}$ & 58.41 & 41.89 & 72.82 & 46.82 & 83.91 & 50.30 \\
& $\mathcal{L}_{doc2docid}+\mathcal{L}_{causal}$ & 59.37 & 42.84 & 73.79 & 47.71 & 84.80 & 50.66 \\
& $\mathcal{L}_{doc2docid}+\mathcal{L}_{causal\_SIDs}$ & 59.14 & 42.61 & 73.58 & 47.51 & 84.65 & 50.48 \\
& SMGR & \textbf{60.10$^{*}$} & \textbf{43.59$^{*}$} & \textbf{74.5$^{*}$} & \textbf{48.23$^{*}$} & \textbf{85.48$^{*}$} & \textbf{51.02$^{*}$} \\
\bottomrule
\end{tabular}
}
\end{table}

\subsection{Online A/B Test (RQ5)}
To evaluate the real-world performance of our approach, we conduct a one-week A/B test on the Meituan platform, utilizing 10\% of the total traffic. Standard industry metrics, including Revenue, CTR, Click, and cost per click (CPC), are used for assessment. All these metrics are the larger the better. As shown in Table ~\ref{tab:online}, compared to the baseline Qwen3-DualTower, our method demonstrates significant improvements across all metrics. These results validate both the deployability and superior effectiveness of our approach in practical scenarios.

\begin{table}[ht]
  \caption{Improvements in the week-long online A/B
test.}
  \label{tab:online}
  \begin{tabular}{ccccc}
    \toprule
    \textbf{Revenue}&\textbf{CTR}&\textbf{Click}&\textbf{CPC}\\
    \midrule
    +1.12\% & +1.02\% & +0.72\% & +0.19\% \\
  \bottomrule
\end{tabular}
\end{table}

\subsection{Case Study (RQ6)}
\begin{figure}[ht]
  \centering
  \includegraphics[width=\columnwidth,height=0.26\textheight]{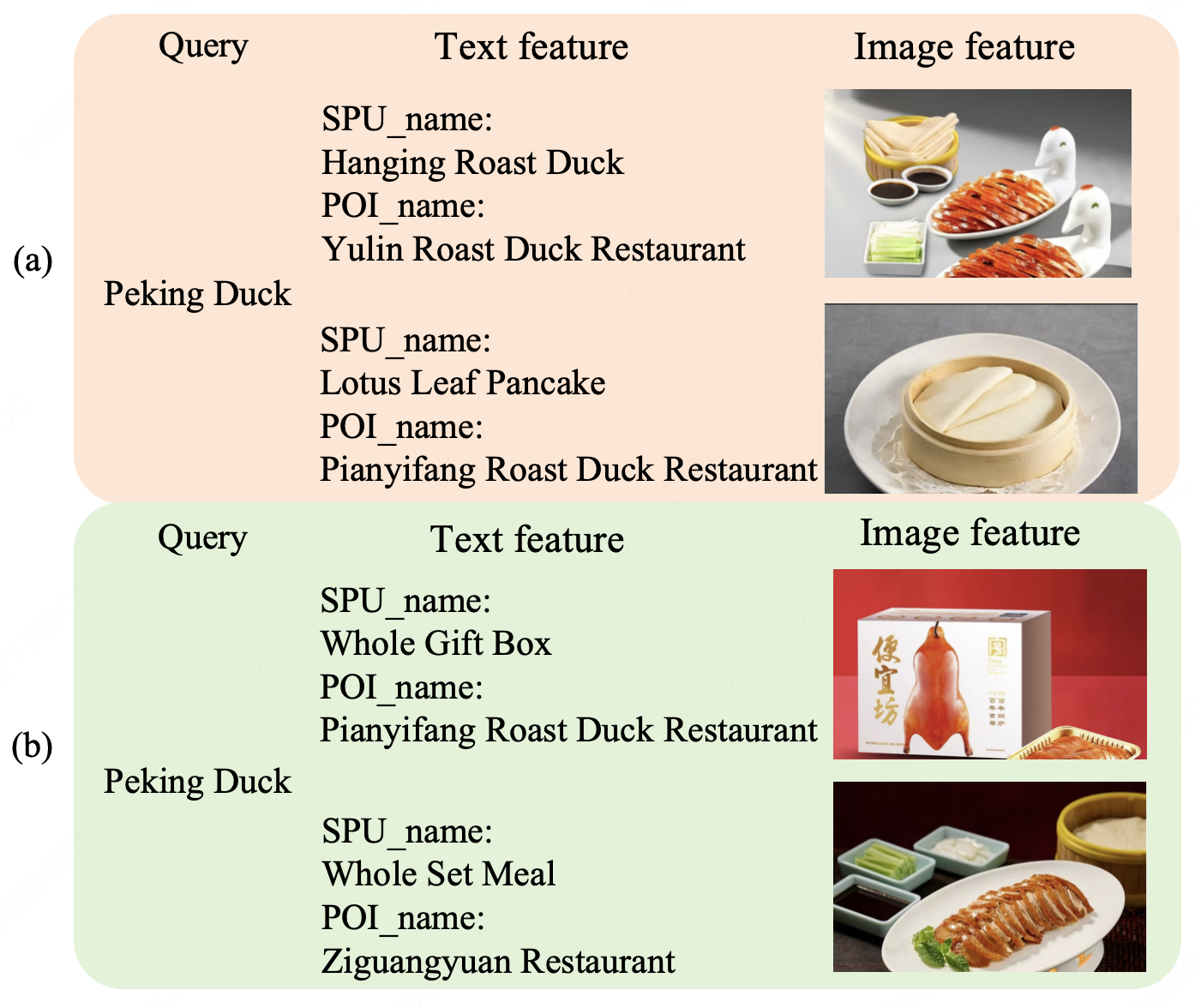}
  \caption{The Top-2 retrieved items for the query "Peking Duck" produced by different models: (a) baseline Joint-Que2search; (b) our proposed SMGR.}
  \Description{The top-2 retrieval items for the query "Peking Duck" produced by different models: (a) baseline Joint-Que2search; (b) our proposed SMGR.}
  \label{fig:case}
\end{figure}

As illustrated in Figure~\ref{fig:case} (a), for the query "Peking Duck", the Joint-Que2search model incorrectly retrieves the item "Lotus Leaf Pancake". This error primarily arises because the POI\_name feature contains the term "Roast Duck", which is semantically related to the query and leads to a false positive retrieval. This case demonstrates that the joint optimization process tends to focus predominantly on textual features while neglecting image features.

In contrast, for the same query "Peking Duck", even when neither the POI\_name ("Ziguangyuan Restaurant") nor the SPU\_name ("Whole Set Meal") contains features directly related to the query, the SMGR model is still able to correctly retrieve the target item by leveraging the image of "Roast Duck". This demonstrates that SMGR can effectively utilize multimodal features to enhance retrieval performance and better respond to user queries.

\section{Conclusion}
In this paper, we identify issues in multimodal retrieval, such as the neglect of modality features and the one-epoch problem caused by joint optimization, and proposes the SMGR method to address these challenges. The staged pretraining strategy in SMGR enables the model to focus on the specific task of each stage, thereby mitigating these problems. By converting high-dimensional embeddings into SIDs, the method alleviates resource and deployment burdens. Furthermore, by introducing both generative and discriminative SIDs understanding and adaptation tasks, our approach improves downstream retrieval performance. Extensive online and offline experiments in real-world scenarios demonstrate the effectiveness of the proposed framework. For future work, we plan to incorporate user historical information on the query side to better capture user preferences. In addition, we intend to explore the generation and utilization of SIDs for sequential data, further broadening the applicability of our approach to more diverse scenarios.

\bibliographystyle{ACM-Reference-Format}
\bibliography{sample-base}

\appendix

\section{Details of RQ-VAE}
\label{appendix:RQ-VAE}

RQ-VAE is a generative model designed to compress high-dimensional continuous embeddings into compact and semantically meaningful discrete representations via multi-layer residual quantization. At each quantization layer $l$, the process is defined as:
\begin{equation}
e^{(l)} = \mathrm{Quantize}^{(l)}(r^{(l)}), \quad r^{(1)} = h, \quad r^{(l+1)} = r^{(l)} - e^{(l)},
\label{eq:rqvae_quantization}
\end{equation}
where $h$ denotes the high-dimensional input embedding, $\mathrm{Quantize}^{(l)}(\cdot)$ denotes the nearest neighbor lookup in the $l$-th codebook, and $L$ is the number of quantization layers. The final discrete semantic representation is formed by concatenating the codebook indices from each layer, $[c^{(1)}, c^{(2)}, \ldots, c^{(L)}]$, which serves as the SIDs and is used as side information in subsequent item tower features.

During RQ-VAE training, the loss function incorporates three components: reconstruction loss, codebook loss, and commitment loss. For an input embedding $h$ and its reconstruction $\hat{h}$, the total loss is given by:
\begin{equation}
\mathcal{L}_{\text{RQ-VAE}} = \| h - \hat{h} \|_2^2 + \beta \sum_{l=1}^{L} \left( \| \mathrm{sg}[r^{(l)}] - e^{(l)} \|_2^2 + \gamma \| r^{(l)} - \mathrm{sg}[e^{(l)}] \|_2^2 \right)
\label{eq:rqvae_loss}
\end{equation}
where the first term, $\| h - \hat{h} \|_2^2$, is the reconstruction loss, measuring how well the encoded discrete SIDs can recover the original embedding and ensuring semantic integrity. The second term, $\| \mathrm{sg}[r^{(l)}] - e^{(l)} \|_2^2$, is the codebook loss, encouraging the encoder outputs to approximate the discrete vectors in the quantization codebook. The third term, $\| r^{(l)} - \mathrm{sg}[e^{(l)}] \|_2^2$, is the commitment loss, which constrains the encoder outputs to adapt to the codebook, promoting collaborative updates between the codebook and feature distribution. Here, $\mathrm{sg}[\cdot]$ denotes the stop-gradient operator, and $\beta$, $\gamma$ are hyperparameters balancing the loss terms.

\section{Hyperparameter Settings}
\label{appendix:hyperparameter}
To ensure consistency in training and evaluation, we employ Qwen3-0.6B as the text encoder for both query and item textual features in all experiments, and utilize a staged pretraining cn-CLIP-ViT-h-14 as the image encoder for item images. Prior to RQ-VAE training, codebooks are initialized using K-means, followed by independent three-layer codebook training for item, image, and text embeddings, with each layer containing 32 entries and no parameter sharing. The item codebook uses 128-dimensional input/output with three hidden layers [128, 64, 32], while the image and text codebooks use 1024-dimensional input/output with four hidden layers [1024, 768, 512, 256]. The commitment loss weight $\beta$ and codebook update loss weight $\gamma$ are both set to 0.25, with candidate values in [0.05, 0.1, 0.15, 0.2, 0.25, 0.3, 0.35, 0.4, 0.45, 0.5].

All experiments are conducted on eight Nvidia A100 GPUs (80GB each), with memory usage ranging from 100GB to 800GB depending on task requirements. The batch size is set to 8 during staged pretraining and 16 during downstream fine-tuning, with gradient accumulation every 8 steps. The temperature for contrastive learning is set to 0.05, with candidate values in [0.01, 0.02, 0.03, 0.04, 0.05, 0.06, 0.07, 0.08, 0.09, 0.1].

\section{Evaluation Protocol}
\label{appendix:evaluation}
For evaluation, both candidate items and queries from the click logs are embedded separately, and FAISS is used to partition the search space to accelerate recall computation. Notably, for a given \texttt{geo\_hash}, queries are only matched with candidate items from the same \texttt{geo\_hash}, reflecting real user behavior. To ensure the reliability of experimental results, each experiment is repeated ten times. A paired t-test is performed to assess the statistical significance of performance differences between models. Results with $p < 0.05$ are considered statistically significant improvements and are marked with $^{*}$ in the tables. Recall@K and NDCG@K are reported as the main evaluation metrics, where $K \in \{5, 10, 20\}$.

\begin{figure*}[ht]
  \centering
  \includegraphics[width=0.8\textwidth,height=0.35\textheight]{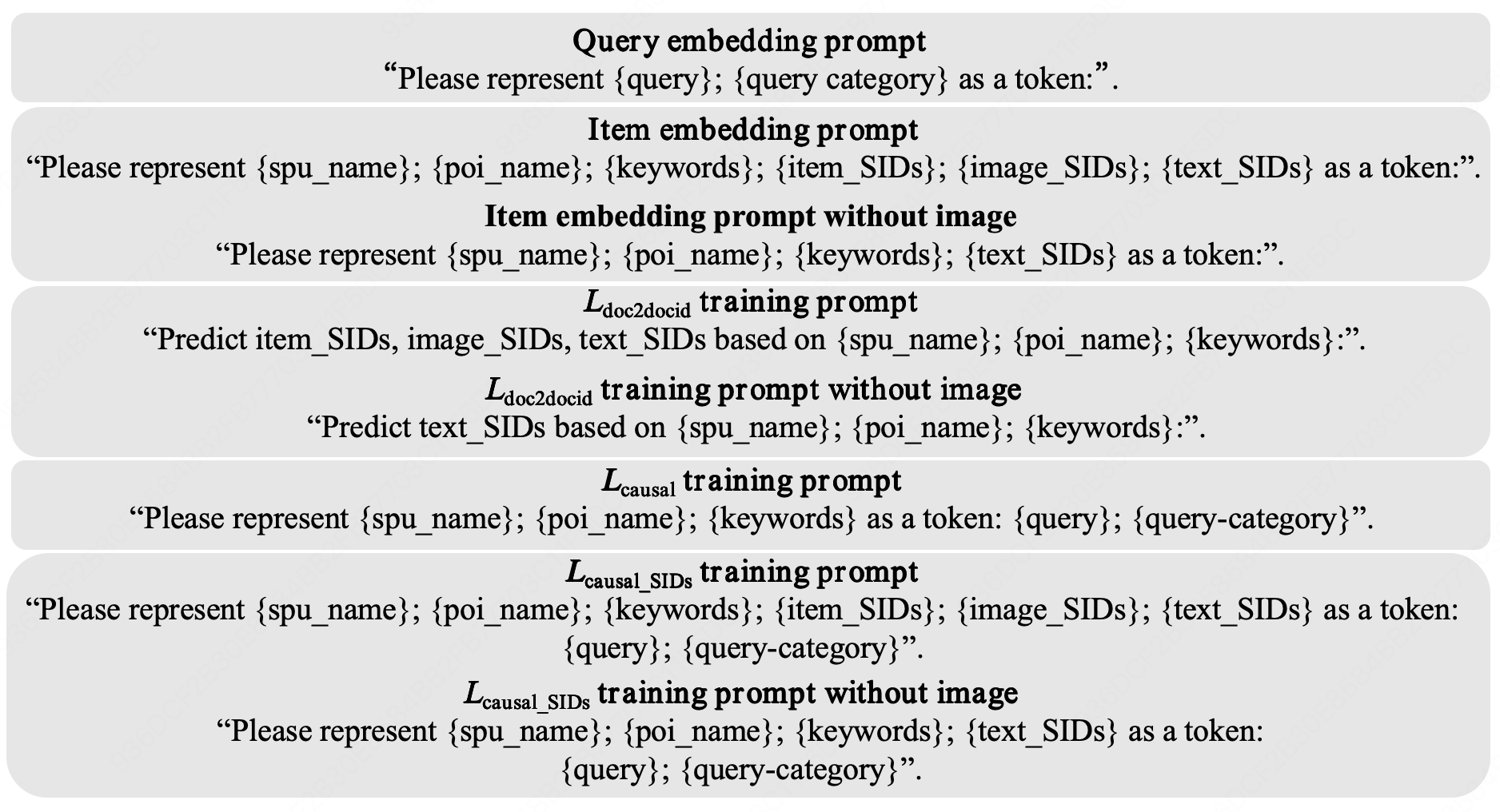}
  \caption{Examples of prompt templates used during training, fine-tuning, and inference.}
  \Description{Examples of prompt templates used during training, fine-tuning, and inference.}
  \label{fig:prompt}
\end{figure*}

\section{Ablation Variants}
\label{appendix:ablation_variants}
\begin{itemize}
    \item \textbf{Joint}: All four losses (image2text, query2image, query2text, query2item) are summed and optimized jointly.
    \item \textbf{Random}: Text features are embedded using the trained Joint model’s text encoder, while image features are replaced with randomly generated vectors.
    \item \textbf{Order1--Order6}: Different staged pretraining orders, including:
    \begin{itemize}
        \item Order1: query2image, image2text, query2text, query2item
        \item Order2: query2image, query2text, image2text, query2item
        \item Order3: query2text, query2image, image2text, query2item
        \item Order4: image2text, query2image, query2text, query2item
        \item Order5: image2text, query2text, query2image, query2item
        \item Order6: query2text, image2text, query2image, query2item
    \end{itemize}
    \item \textbf{Item-only}: Uses only item\_SIDs as item-side information.
    \item \textbf{Image-text}: Uses image\_SIDs and text\_SIDs as item-side information.
    \item \textbf{Item-image}: Uses image\_SIDs and item\_SIDs as item-side information.
    \item \textbf{Item-text}: Uses text\_SIDs and item\_SIDs as item-side information.
    \item \textbf{All\_SIDs}: Uses image\_SIDs, text\_SIDs, and item\_SIDs as item-side information.
    \item $\mathcal{L}_{\text{doc2docid}}$: Only performs SIDs prediction for trained models.
    \item $\mathcal{L}_{\text{doc2docid}} + \mathcal{L}_{\text{causal}}$: Performs both SIDs prediction and causal prediction fine-tuning on sentences containing item-side (without SIDs) and query-side features.
    \item $\mathcal{L}_{\text{doc2docid}} + \mathcal{L}_{\text{causal\_SIDs}}$: Performs both SIDs prediction and causal prediction fine-tuning on sentences containing item-side (with SIDs) and query-side features.
\end{itemize}

\section{Prompt Templates}
\label{appendix:prompt}
As illustrated in Figure~\ref{fig:prompt}, representative examples of prompt templates utilized during model training, fine-tuning, and inference are presented.
\end{document}